# Laboratory Gas-phase Infrared Spectra of Two Astronomically Relevant PAH Cations: Diindenoperylene, $C_{32}H_{16}^+$ and Dicoronylene, $C_{48}H_{20}^+$

Junfeng Zhen[1,2,3,4], Alessandra Candian[1], Pablo Castellanos[1,2], Jordy Bouwman[2,5],
Harold Linnartz[2], and Alexander G. G. M. Tielens[1]
[1] Leiden Observatory, Leiden University, P.O. Box 9513, 2300 RA Leiden, The Netherlands; jfzhen@ustc.edu.cn
[2] Sackler Laboratory for Astrophysics, Leiden Observatory, Leiden University, P.O. Box 9513, 2300 RA Leiden, The Netherlands
[3] CAS Key Laboratory for Research in Galaxies and Cosmology, Department of Astronomy, University of Science and Technology of China,
Hefei 230026, People's Republic of China
[4] School of Astronomy and Space Science, University of Science and Technology of China, Hefei 230026, People's Republic of China
[5] Radboud University, Institute for Molecules and Materials, FELIX Laboratory, Toernooiveld 7c, 6525ED Nijmegen, The Netherlands

## Abstract

The first gas-phase infrared spectra of two isolated astronomically relevant and large polycyclic aromatic hydrocarbon (PAH) cations—diindenoperylene (DIP) and dicoronylene (DC)—in the 530–1800 cm$^{-1}$ (18.9–5.6 $\mu$m) range—are presented. Vibrational band positions are determined for comparison to the aromatic infrared bands. The spectra are obtained via infrared multiphoton dissociation spectroscopy of ions stored in a quadrupole ion trap using the intense and tunable radiation of the free electron laser for infrared experiments (FELIX). DIP$^+$ shows its main absorption peaks at 737 (13.57), 800 (12.50), 1001 (9.99), 1070 (9.35), 1115 (8.97), 1152 (8.68), 1278 (7.83), 1420 (7.04), and 1550 (6.45) cm$^{-1}$ ($\mu$m), in good agreement with density functional theory (DFT) calculations that are uniformly scaled to take anharmonicities into account. DC$^+$ has its main absorption peaks at 853 (11.72), 876 (11.42), 1032 (9.69), 1168 (8.56), 1300 (7.69), 1427 (7.01), and 1566 (6.39) cm$^{-1}$ ($\mu$m), which also agree well with the scaled DFT results presented here. The DIP$^+$ and DC$^+$ spectra are compared with the prominent infrared features observed toward NGC 7023. This results both in matches and clear deviations. Moreover, in the 11.0–14.0 $\mu$m region, specific bands can be linked to CH out-of-plane (oop) bending modes of different CH edge structures in large PAHs. The molecular origin of these findings and their astronomical relevance are discussed.

## 1. Introduction

Strong emission features at 3.3, 6.2, 7.7, 8.6, and 11.2 $\mu$m dominate the infrared (IR) spectrum of many astronomical sources. These bands are commonly known as the aromatic infrared bands (AIBs) and are generally attributed to IR fluorescence of large (roughly more than 40 C atoms containing) polycyclic aromatic hydrocarbon (PAH) molecules and their related families. These bands are emitted upon ultraviolet (UV) excitation of these species (Sellgren 1984; Allamandola et al. 1989; Puget & Leger 1989). PAHs are found to be abundant and ubiquitous, and they are expected to account for ∼10% of the cosmic carbon (Tielens 2008). They play an important role in the energy and ionization balance of the interstellar medium (ISM) and may serve as a catalyst for the formation of molecular H$_2$ in photo-dissociation regions (PDRs; Tielens 2013, and references therein).

The AIBs have been interpreted as the cumulative spectrum of a family of PAHs and PAH cations as well as PAH derivatives (Tielens 2008, and references therein). For this reason, a variety of neutral and ionized PAH molecules with different sizes and structures have been studied, both by theory and in the laboratory, to link specific (sets of) AIB features to (modes of specific) carriers (Sloan et al. 1999; Hony et al. 2001; Malloci et al. 2004; Pathak & Rastogi 2008, for example). Experimentally, IR emission gas-phase spectra are available from studies by Cook et al. (1998), Kim & Saykally (2002). Infrared absorption spectra of PAHs and PAH cations have been recorded in rare gas matrices (e.g., Hudgins & Allamandola 1995; Mattioda et al. 2003; Bernstein et al. 2007; Tsuge et al. 2016). In the gas phase, spectra have been measured using infrared multiphoton dissociation (IRMPD; Oomens et al. 2001), messenger atom photo-dissociation spectroscopy (Piest et al. 1999; Ricks et al. 2009), and ion-dip technique (Maltseva et al. 2015, 2016). The majority of the available studies have focused on smaller PAHs, like naphthalene and coronene that are commercially available and relatively easy to handle in a laboratory setting. Spectra of large and astronomically more relevant PAHs—from ∼40 C atoms upwards—are much rarer (Kokkin et al. 2008; Zhen et al. 2016). Recently, the first gas-phase IR spectra of a large PAH cation—HBC or hexa-peri-hexabenzocoronene (C$_{42}$H$_{18}$)—and its dication have been reported by Zhen et al. (2017).

Tielens (2013) first proposed the concept of "grandPAHs," as a set of the chemically most stable PAH species that are able to survive in the harsh conditions of the ISM. This idea was motivated by the observation of highly similar AIB spectra observed toward very different interstellar sources as well as the limited number of bands in the 15–20 $\mu$m range, which is a region in which structural features of PAHs are expected to show up (Boersma et al. 2010). The "grandPAHs" concept is currently under further investigation (e.g., Andrews et al. 2015; Peeters et al. 2017). If "grandPAHs" indeed abundantly exist in the ISM, this would simplify the picture for the scientific community, as only a rather limited number of species would be important to characterize (Tielens 2013; Andrews et al. 2015; Croiset et al. 2016). The present study adds to this discussion by presenting the gas-phase IRMPD spectra of two other large PAH cations—diindenoperylene (DIP, $C_{32}H_{16}$, $m/z = 400.13$) and dicoronylene (DC, $C_{48}H_{20}$, $m/z = 596.16$). The molecular geometries are shown as insets in Figure 1. These spectra are compared to both density functional theory (DFT) predictions and astronomical spectra.

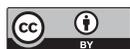

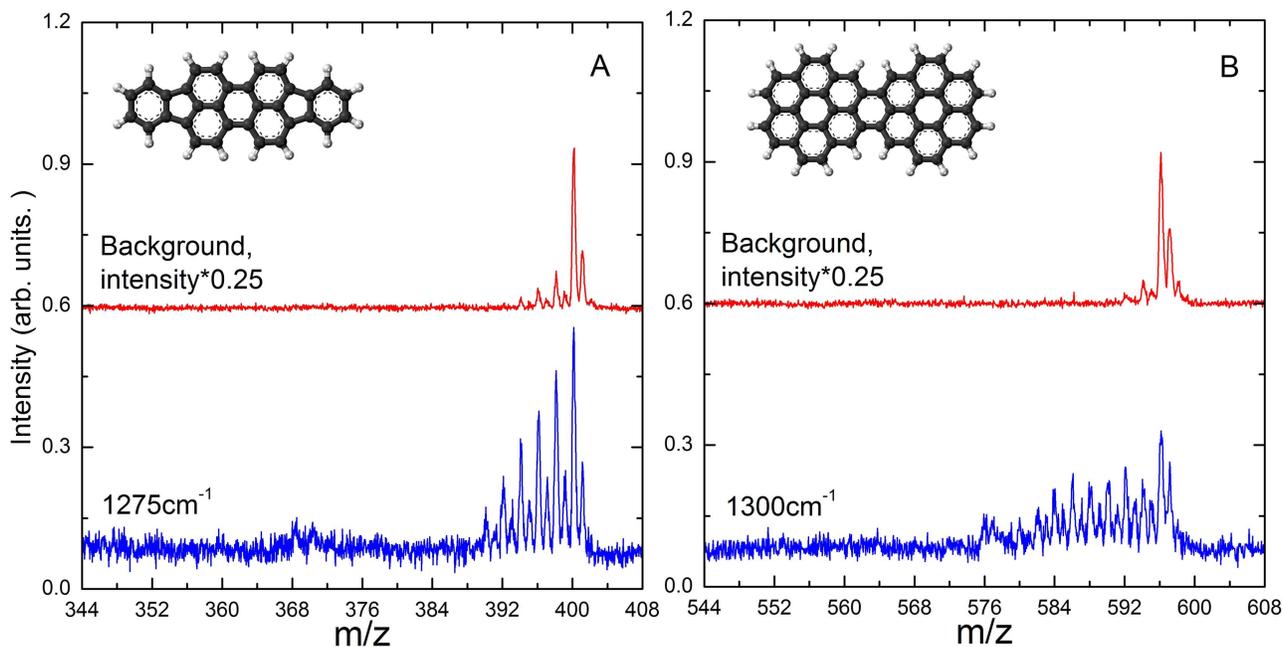

**Figure 1.** Mass spectrum showing the electron impact (upper graph) and photo-induced (lower graph) products resulting from the irradiation of DIP$^+$ at 1275 cm$^{-1}$ (panel A) and DC$^+$ at 1300 cm$^{-1}$ (panel B).

## 2. Experimental and Theoretical Methods

The experiments described here have been performed with iPoP, our fully mobile "instrument for Photo-dissociation of PAHs" (Zhen et al. 2014b). The set-up has been used at one of the end stations of FELIX, the Free Electron Laser for InfraRed eXperiments at Radboud University (Oepts et al. 1995). The information on the experimental procedures is available from (Zhen et al. 2017). Here, only the relevant details are provided.

The two central parts of iPoP are a quadrupole ion trap (QIT) and a time-of-flight (TOF) mass spectrometer. Gas-phase DIP and DC precursor molecules are obtained by heating commercially available powder (Kentax, purity higher than 99.5%) in an oven that can be preset to a value very close to the sublimation temperature: ∼480 K for DIP and ∼670 K for DC. The gas-phase neutrals are ionized using electron impact ionization, typically with 83 eV electron impact energy. The resulting DIP$^+$ and DC$^+$ cations enter in the QIT via an ion gate and are trapped by applying a 1 MHz radiofrequency electric field (3000 and 3280 $V_{p-p}$) onto the ring electrode. Helium that is continuously introduced into the ion trap thermalizes the ion cloud.

Spectra between 530 and 1800 cm$^{-1}$ (18.9–5.6 $\mu$m) are obtained by means of IRMPD, using the intense and tunable radiation of FELIX. This free electron laser delivers 5 $\mu$s long macro-pulses of light with an energy up to 100 mJ and at a repetition rate of 10 Hz. Its bandwidth (full-width-at-half-maximum, FWHM) amounts to about 0.6% of the central wavelength (i.e., ranging from 3 to 10 cm$^{-1}$). Resonant vibrational excitation through multiple photon absorptions causes the PAH cation to fragment. Subsequently, the fragments are extracted from the QIT and analyzed mass spectrometrically in a TOF mass spectrometer. All (stronger) channels resulting from the photo-fragmentation of the parent are summed up and this value is normalized to the total signal, i.e., the value found for parent plus photo-fragment ions. This results in a relative photo-fragmentation intensity. The IR spectrum of the trapped ion is then obtained by recording the fragment ion yield as a function of the wavelength.

The ion cloud is irradiated for 0.8 s (8 macro-pulses) to obtain a single IRMPD mass spectrum. The average of 25 individual mass spectra is taken for a data point at a single wavelength. A spectrum is obtained by averaging mass spectra while tuning the free electron laser with 5 cm$^{-1}$ steps. This value is close to the absolute wavelength accuracy of the system. In addition, blank spectra are recorded in order to subtract fragmentation signals that are present in absence of IR radiation, i.e., upon electron impact ionization only.

The recorded spectra are normalized to the pulse energy, which, in the best case, offers only a first-order correction, as IRMPD is a nonlinear process. Signal intensities, therefore, should be handled with care, as these reflect more the dissociation efficiency upon excitation of a specific vibrational mode than its IR band strength. Moreover, as high excitations are involved in the detection process, anharmonic effects may shift absorption wavelengths by as much as several cm$^{-1}$ (Oomens et al. 2001). A direct comparison with astronomical data, therefore should be done with care: intensities may vary and wavelength positions may be somewhat off-set. For an in-depth comparison, detailed theoretical studies on the effects of anharmonicity in highly excited PAHs (Parneix et al. 2013) are required, which is currently outside of the realm for large PAHs.

To further guide the spectral interpretation of the experimental spectra, DFT calculations are performed with the B3LYP functional and the 6–31 G** basis set, using Gaussian 09. It has been shown that multiple scaling factors are needed when a larger basis set than the standard 4-31G is used (Langhoff 1996; Buragohain et al. 2015). In this work, a uniform scaling of 0.961 is applied to all modes, which suffices to unambiguously assign the broad bands in the IRMPD spectra. For both DIP$^+$ and DC$^+$, only the doublet state is considered, as this state is lowest in energy (Bauschlicher & Bakes 2010a; Bauschlicher et al. 2010b).

### 3. Results and Discussion

Figure 1 shows the mass spectra for DIP$^+$ (Figure 1(A)) and DC$^+$ (Figure 1(B)) for fragments formed upon electron impact ionization only (upper graphs) and additional photo-excitation (lower graphs) using 1275 cm$^{-1}$ photons (DIP$^+$) and 1300 cm$^{-1}$ photons (DC$^+$). Due to the impacting electrons, the mass spectra before irradiation (labeled as background in the upper graphs) exhibit a small amount of residual fragmentation. The mass signals also comprise small contributions from fragments containing $^{13}$C isotopes. The mass spectrum in Figure 1(A) shows that the DIP$^+$ photo-dissociation pattern preferably follows sequential H$_2$ (or 2H) loss channels, similarly to what was observed for HBC$^+$ by Zhen et al. (2014b), but the hydrogen stripping is not complete and no bare C$_{32}^+$ ($m/z = 384$) signal is observed. In Figure 1(A), also a (much weaker) carbon loss channel is clear found, resulting in noisy signals around the masses of (C$_{30}$H$_8^+$, $m/z = 368$). Figure 1(B) shows the mass spectrum of DC$^+$. The photo-induced fragmentation pattern visualizes a more complete dehydrogenation process, compared to DIP$^+$, and even the fully dehydrogenated, i.e., bare carbon (C$_{48}^+$, $m/z = 576$) geometry is observed. No detectable evidence is found for a loss channel involving carbon.

As pointed out by earlier studies (Ekern et al. 1998; Zhen et al. 2014a), large PAH cations will initially fragment through rapid H-loss, leaving a bare carbon skeleton. The involved fragmentation kinetics is controlled by the molecular absorption properties of the parent species and the competition of the different processes involved; e.g., their relative rates, which are set by the energy barriers and the tightness of the transition states involved (Tielens 2008). This competition between the different fragmentation channels is known to depend on size as, in contrast to large PAHs, small PAHs already start to lose C$_2$ units before all H atoms are lost (Ekern et al. 1998; West et al. 2014). Molecular geometry may also affect the fragmentation behavior as the binding energy per C-atom is higher for compact PAHs than for non-compact PAHs (Ricca et al. 2012). Our spectra show differences in fragmentation behavior for these two species with DIP$^+$ losing some C$_2$ before H-loss is complete while DC$^+$ does not (Figure 1). Whether this reflects the difference in size or molecular geometry remains to be determined. We do conclude, though, that IRMPD provides an alternative way to study the fragmentation patterns of highly excited PAHs. In space, PAH evolution will be a competition between UV-driven fragmentation and reactions with H atoms (Vuong & Foing 2000; Le Page et al. 2003). Further IRMPD studies may serve as inputs for astronomical models for the evolution of interstellar PAHs (Andrews et al. 2015; Berné et al. 2015).

In Figure 2, the resulting IR spectra for DIP$^+$ (panel A) and DC$^+$ (panel B) between 530 and 1800 cm$^{-1}$ are presented. The molecular geometries are repeated for convenience. The two spectra exhibit a number of clearly resolved vibrational bands with reasonable signal-to-noise ratios. Both relatively narrow and broader bands are found. Broadening is caused by overlapping bands or may be due to anharmonic effects. This becomes clear from the scaled harmonic DFT predictions for the vibrational band positions that are incorporated as sticks in the figure. Table 1 summarizes all vibrational and computed wavelengths. Generally, one can see a quite convincing agreement between experiment and theory. The vibrational assignment of the experimental bands, therefore, is based on the theoretical predictions. A comparison of the involved intensities is less obvious, as stated before, because the IRMPD intensities do not directly reflect absorption strengths. This is discussed in more detail below.

In Figure 2(A) and Table 1, a comparison of our experimental IRMPD DIP$^+$ spectrum with the theoretical predictions is presented. Assignments can be made that are based on a smallest wavelength difference criterion. This also works for bands that are very close, resolved in the calculations, but that are not resolved in the measured spectra. For example, the broad experimental band around 1278 cm$^{-1}$ overlaps with theoretically predicted stronger bands at 1232, 1308, and 1333 cm$^{-1}$. It is very likely that the 1278 cm$^{-1}$ band also includes bands at 1248, 1287, and 1313 cm$^{-1}$. This explains why the band is both intense and broad with a FWHM of around 120 cm$^{-1}$. Figure 2(B) and Table 1 provide similar information for DC$^+$. Also the DC$^+$ experimental and theoretical results are rather similar and the recorded bands are assigned by looking for (close) matches with the predicted ones. For the theoretical result of DC$^+$, we note that in the out-of-plane (oop) modes region (11–14 $\mu$m), DC$^+$ shows a relatively strong "solo" mode at 870 cm$^{-1}$ (compare to the longer wavelengths oop modes) despite that this species has only 4 "solo" Hs as compared to 16 "duo" Hs; in contrast, in Figure 2(B), we note that the "duo" oop mode in DC$^+$ at 853 cm$^{-1}$ seems to be the same strong comparison to the "solo" oop mode at 876 cm$^{-1}$. This may indicate that the nonlinear aspect of the IRMPD technique seems to play a major role in intensity prediction rather than the anharmonicity effect, as the experimental "solo" modes are as strong as the "duo" modes.

A comparison between Figures 2(A) and (B) shows clear differences. DIP$^+$ exhibits multiple CH in-plane bending vibration bands (1001, 1070, 1115, and 1152 cm$^{-1}$) in a range (1000–1200 cm$^{-1}$) where DC$^+$, and also HBC$^+$ show only two resolved bands (Zhen et al. 2017). The origin of these features can be traced back to a combination of CH in plane and CC stretches modes involving also, but not exclusively, the two pentagons in the DIP geometry. Thus, if this is confirmed in other PAHs, these bands can offer a tool to identify PAHs with pentagon structures. Another pronounced difference is found around 1600 cm$^{-1}$; whereas HBC$^+$ (Zhen et al. 2017) only exhibits a relatively weak feature in this range, both DIP$^+$ an DC$^+$ have much stronger bands here. Further studies are required to assess whether this reflects the less symmetric structure of the species studied here, which could "activate" more CC modes than for the more highly symmetric HBC$^+$.

As a general remark, we mention that, despite the fact that a scaling factor has been applied to correct the DFT calculations for anharmonicities, most experimentally measured band positions of DIP$^+$ and DC$^+$ are shifted by several wavenumbers with respect to the calculated positions. Similar findings were found in other studies (Oomens et al. 2003). The scaling factors are based on comparison between DFT calculations and low temperature matrix isolation studies (Langhoff 1996). As stated earlier, the IRMPD (absorption) process results in highly excited species and will therefore induce an additional (red) shift. The size of this shift depends on the level of anharmonicity and the temperature of the involved modes (Chen et al. 2016;

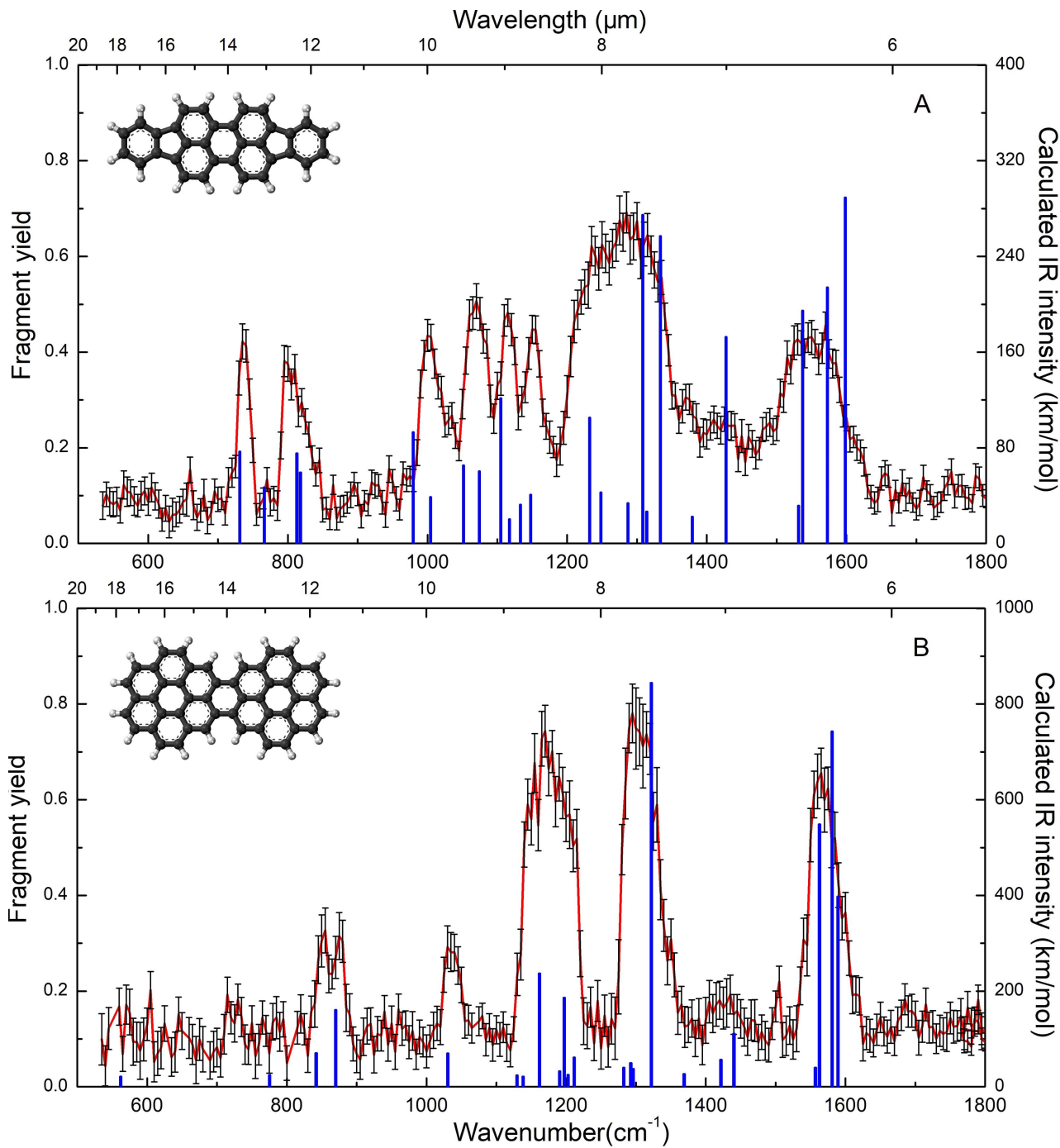

**Figure 2.** Measured IRMPD spectrum of gas-phase DIP$^+$ (upper plot) and DC$^+$ (lower plot). The computed vibrational peak positions scaled to correct for anharmonicities are represented by vertical bars.

Candian & Mackie 2017). With this qualifying remark in mind, the comparison of our experimental results with the theoretical predictions is very encouraging with deviations of the order of a few percent.

## 4. Astrophysical Relevance

The IRMPD experiments described here, in combination with the theoretical predictions, yield band positions of vibrational bands that can be compared with astronomical

Table 1
Infrared Transitions Measured and Calculated for DIP$^+$ and DC$^+$

| DIP$^+$ | | | DC$^+$ | | |
|---|---|---|---|---|---|
| Experimental Band cm$^{-1}$, μm | Computed Band[a] cm$^{-1}$, km mol$^{-1}$ | Mode Description | Experimental Band cm$^{-1}$, μm | Computed Band[a] cm$^{-1}$, km mol$^{-1}$ | Mode Description |
| 737 (s[b], A[c], 15[d]), 13.57[e] | 731, 77[f] | oop CH bending | | | |
| | 766, 47 | | | 562, 21 | |
| 800 (s, A, 30), 12.50 | 813, 75 | oop CH bending | | 775, 24 | |
| | 818, 59 | | 853 (m, A, 15), 11.72 | 842, 70 | oop CH bending |
| | 979, 93 | | 876 (m, A, 15), 11.42 | 870, 160 | oop CH bending |
| 1001 (s, A, 30), 9.99 | 1004, 39 | in-plane CH bending | 1032 (m, A, 20), 9.69 | 1030, 69 | in-plane CH bending |
| | 1052, 65 | | | 1130, 23 | |
| 1070 (s, A, 20), 9.35 | 1074, 60 | in-plane CH bending | | 1138, 21 | |
| | 1105, 121 | | | 1161, 236 | |
| 1115 (s, A, 30), 8.97 | 1117, 20 | in-plane CH bending | | 1190, 32 | |
| | 1133, 32 | | 1168 (s, B, 60), 8.56 | 1197, 186 | in-plane CH bending |
| 1152 (s, A, 20), 8.68 | 1148, 41 | in-plane CH bending | | 1202, 25 | |
| | 1232, 105 | | | 1211, 61 | |
| | 1248, 43 | | | 1282, 39 | |
| | 1287, 34 | | | 1292, 49 | |
| 1278 (s, B, 120), 7.83 | 1308, 275 | CC stretching | | 1296, 37 | |
| | 1313, 27 | | 1300 (s, B, 40), 7.69 | 1322, 844 | CC stretching |
| | 1333, 257 | | | 1369, 26 | |
| | 1379, 22 | | | 1421, 56 | |
| 1420 (m, C, 50), 7.04 | 1427, 173 | CC stretching | 1427 (w, C, 40), 7.01 | 1440, 110 | CC stretching |
| | 1531, 31 | | | 1556, 40 | |
| | 1537, 195 | | | 1562, 548 | |
| 1550 (s, B, 80), 6.45 | 1572, 214 | CC stretching | 1566 (s, B, 40), 6.39 | 1580, 742 | CC stretching |
| | 1598, 289 | | | 1589, 397 | |

**Notes.**
[a] Theoretical band positions have been corrected by a scaling factor of 0.961 to account for anharmonicity.
[b] Relative intensities are indicated as w, m, s for weak, medium, strong.
[c] Experimental band position uncertainties are estimated to be A (1 ∼ 3 cm$^{-1}$), B (3 ∼ 6 cm$^{-1}$) and C (15 ∼ 30 cm$^{-1}$).
[d] FWHM (cm$^{-1}$).
[e] We give approximate values.
[f] Theoretical intensities are presented, only transitions with intensities larger than 20 km mol$^{-1}$ are listed.

data. The laboratory band positions suffer to some extent from anharmonic effects that, however, can be taken into account, also because the observed bands are generally rather broad, of the order of 20−40 cm$^{-1}$, comparable to typical AIB values (Tielens 2008; Bauschlicher & Bakes 2010a; Bauschlicher et al. 2010b). However, the laboratory band intensity values have to be used with care, as—unlike direct absorption spectroscopy—IRMPD is based on a nonlinear process. We also note that observed AIB intensities will be affected by the excitation temperature of the emitting interstellar PAHs, which can be quite different from those in our laboratory study (Verstraete et al. 2001; Boersma et al. 2011; Chen et al. 2016; Candian & Mackie 2017).

In Figure 3, the obtained experimental IRMPD spectra for DIP$^+$ and DC$^+$ are replotted together with the previously reported spectrum of HBC$^+$ (Zhen et al. 2017). Also included in the figure (top row) is a typical AIB emission spectrum as observed with Spitzer toward NGC 7023 (Houck et al. 2004; Werner et al. 2004). The DC$^+$ spectrum exhibits clear resemblances with the AIB spectrum; the strong bands at 6.2, 7.7, 8.6, and 9.7 μm closely match the position (and even intensity profile) observed in the AIB spectrum. The spectral patterns of DIP$^+$ and HBC$^+$ show much less agreement, even when taking into account that the laboratory intensities do not need to match.

In the geometries shown in Figure 3, the numbers of adjacent CH groups per aromatic ring are given.[6] Experimental and theoretical studies on small PAHs have shown that the pattern of the oop CH bending modes is very characteristic for the number of adjacent CH groups (Hony et al. 2001), and quantum chemical studies have extended this finding to large PAHs (Boersma et al. 2014). The number of adjacent CH groups clearly differs for the three PAH cations shown in Figure 3. Comparing the spectra to the observed bands in the range of the CH oop bending modes (11.0–14.0 μm) of NGC 7023—which shows very clear (major) emission features at 11.2, 12.0, 12.7, and 13.5 μm—we conclude that the interstellar features are close to several of the observed laboratory bands: the interstellar 11.2 μm band is very close to the 11.42 μm band due to "solo" CH-structure bending modes in DC$^+$; the interstellar 12.0 μm band is not far from the 11.72 μm and 12.50 μm band due to "duo" CH-structure bending modes in DC$^+$ and DIP$^+$; the interstellar 13.1 μm band is close to the 12.7 μm band due to "trio" CH-structure bending modes in HBC$^+$; the interstellar 13.5 $\mu$m band nearly coincides with the 13.57 $\mu$m band due to "quartet" CH-structure bending modes in DIP$^+$.

---

[6] 1 is indicative of aromatic rings carrying CH groups that have no neighboring CH groups (termed "non-adjacent" or "solo" CH groups); likewise, 2 is indicative of two adjacent CH groups ("doubly adjacent" or "duo" CH's); 3 is indicative of three adjacent CH groups ("triply adjacent" or "trio" CH's); and 4 is indicative of four adjacent CH groups ("quadruply adjacent" or "quartet" CH's; Hony et al. 2001).

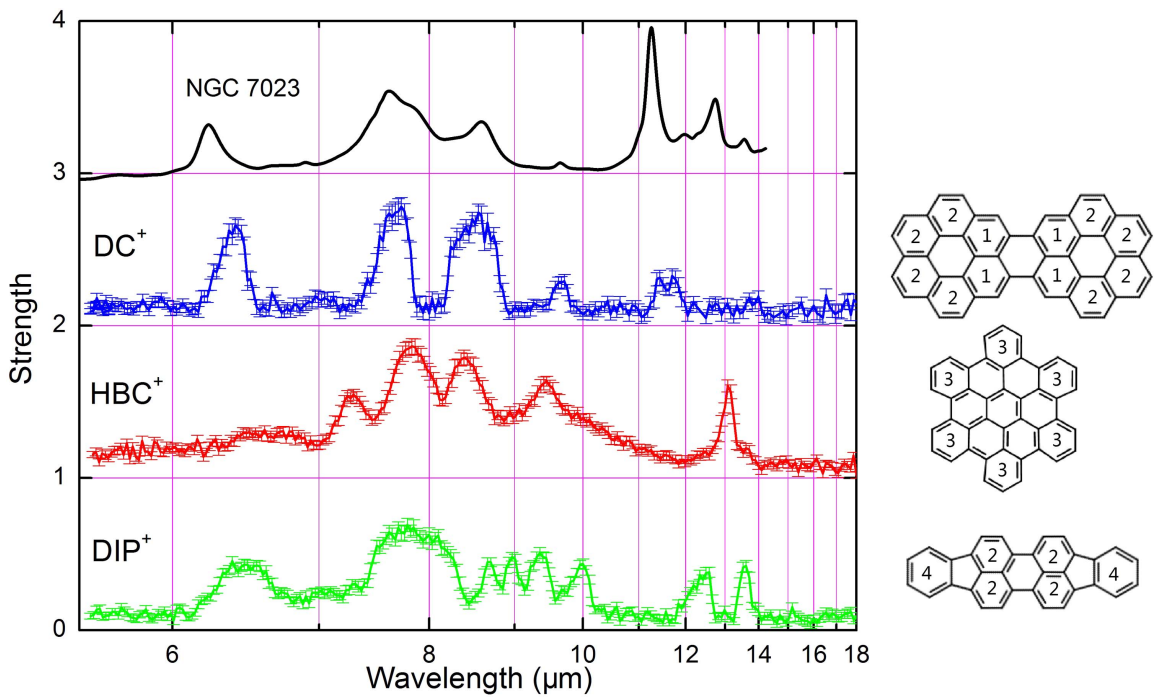

**Figure 3.** IR spectra of DC$^+$, HBC$^+$, and DIP$^+$ (lower three panels) compared to one representative AIB emission spectrum toward NGC 7023 (upper panel). The latter has been recorded using *Spitzer* and the mid-IR spectra were taken with the short-low (SL) module covering the wavelength range 5.1–14.2 $\mu$m (AORs 3871488 and 3871744) (Houck et al. 2004; Werner et al. 2004). The numbers in the molecular geometries, shown on the right, indicate the number of adjacent CH groups per aromatic ring.

As stated earlier, the AIB spectrum is thought to be composed of the individual spectra of a family of different PAHs that will have different absorption efficiencies and that will be present in space in different abundances. A one-to-one comparison of a PAH and the AIBs, for that reason, is unlikely to result in a full reproduction of the astronomical spectrum. Nevertheless, we note that DC$^+$ gets very close, a finding that is in agreement with theoretical studies that show that IR spectra of large compact PAHs provide the better agreement with astronomical observations (Ricca et al. 2012).

Is it possible to extract more general findings from these data? Typically, many different PAHs exhibit more or less similar spectral features, as similar vibrational modes are involved. Nevertheless, differences can be found, as apparent in Figure 3. Even though premature, on the basis of only the few spectra presented here, one could conclude that a larger PAH, compact and possibly better resistant against UV-induced dissociation of its carbon frame, may be a more likely carrier of the AIBs than smaller or less compact PAHs. A preliminary conclusion based on the small data set presented here would be that "GrandPAHs" must be compact (Tielens 2013; Andrews et al. 2015; Croiset et al. 2016). This would make sense, but definitely needs to be tested by investigating other PAHs.

## 5. Conclusions

The IR spectra of DIP$^+$ and DC$^+$, in the 530−1800 cm$^{-1}$ (18.9−5.6 $\mu$m) IR fingerprint region, have been measured and compared to DFT calculated and measured interstellar spectra. The applied IRMPD technique provides useful band positions for IR active modes, despite inherent inaccuracies such as band shifts or nonlinear band intensities. A comparison of the recorded spectra with AIBs supports— but is not conclusive evidence for—the possibility that large compact PAHs act as AIB carriers. The spectra also demonstrate that the general rules for the CH oop bending modes derived for small PAHs also apply to large PAHs and allow us to interpret the AIB bands in the 11.0–14.0 $\mu$m range as being due to CH oop bending modes associated with "solo", "duo", "trio", or "quartet" structures on the PAH edges. Finally, we note that IRMPD also provides a tool for the study of the fragmentation behavior of PAHs and that even large PAHs are amenable to study using a IR laser facility.

The authors gratefully acknowledge the FELIX staff for their technical assistance during preparation and the actual beamline shifts. Studies of interstellar chemistry at Leiden Observatory are supported through advanced-ERC grant 246976 from the European Research Council, through a grant by the Netherlands Organisation for Scientific Research (NWO) as part of the Dutch Astrochemistry Network, and through the Spinoza premie. J.Z. acknowledges financial support from National Science Foundation of China (NSFC, grant No. 11743004). J.B. and A.C. acknowledge NWO for a Veni grant (722.013.014 and 639.041.543, respectively). We acknowledge the European Union (EU) and Horizon 2020 funding awarded under the Marie Sklodowska-Curie action to the EUROPAH consortium, grant No. 722346. DFT calculations were carried out on the Dutch national e-infrastructure Cartesius with the support of SURF Cooperative (grant SH-362-15).

### ORCID iDs

Alessandra Candian  https://orcid.org/0000-0002-5431-4449
Pablo Castellanos  https://orcid.org/0000-0002-8585-8035
Harold Linnartz  https://orcid.org/0000-0002-8322-3538